\documentclass[twocolumn]{article}

\usepackage{amsmath,amssymb}
\usepackage{pstricks}
\usepackage{graphicx}
\usepackage{xspace}            
\usepackage{graphicx}
\usepackage{multirow}
\usepackage{subfig}
\usepackage{multirow}
\usepackage{array}
\usepackage{url}
\usepackage{pdfpages}
\usepackage{booktabs}
\usepackage{balance}

\usepackage{authblk}

\usepackage{url}

\newcommand{\vct}[1]{\ensuremath{\boldsymbol{#1}}}

\newcommand{\set}[1]{\ensuremath{\mathcal{#1}}}
\newcommand{\con}[1]{\ensuremath{\mathsf{#1}}}

\newcommand{\myparagraph}[1]{\smallskip \noindent \textbf{#1}}

\newcommand{\ie}{\emph{i.e.}\xspace}
\newcommand{\eg}{\emph{e.g.}\xspace}
\newcommand{\etal}{\emph{et al.}\xspace}

\newcommand{\deltaphish}{\texttt{$\delta$Phish}\xspace}

\begin{document}

\title{DeltaPhish: Detecting Phishing Webpages\\in Compromised Websites\footnote{Preprint version of the work accepted for publication at ESORICS 2017.}}

\author[1,2]{Igino Corona}
\author[1,2]{Battista Biggio}
\author[2]{Matteo Contini}
\author[1,2]{Luca Piras}
\author[2]{Roberto Corda}
\author[2]{Mauro Mereu}
\author[2]{Guido Mureddu}
\author[1,2]{Davide Ariu}
\author[1,2]{Fabio Roli}

\affil[1]{Pluribus One, via Bellini 9, 09123 Cagliari, Italy}
\affil[2]{DIEE, University of Cagliari, Piazza d'Armi 09123, Cagliari, Italy}
\date{}                     
\setcounter{Maxaffil}{0}

\maketitle

\abstract{The large-scale deployment of modern phishing attacks relies on the automatic exploitation of vulnerable websites in the wild, to maximize profit while hindering attack traceability, detection and blacklisting.
To the best of our knowledge, this is the first work that specifically leverages this adversarial behavior for detection purposes. We show that phishing webpages can be accurately detected by highlighting HTML code and visual differences with respect to other (legitimate) pages hosted within a compromised website.
Our system, named DeltaPhish, can be installed as part of a web application firewall, to detect the presence of anomalous content on a website after compromise, and eventually prevent access to it.
DeltaPhish is also robust against adversarial attempts in which the HTML code of the phishing page is carefully manipulated to evade detection. We empirically evaluate it on more than 5,500 webpages collected in the wild from compromised websites, showing that it is capable of detecting more than 99\% of phishing webpages, while only misclassifying less than 1\% of legitimate pages. We further show that the detection rate remains higher than 70\% even under very sophisticated attacks carefully designed to evade our system.}

\section{Introduction}
In spite of more than a decade of research, phishing is still a concrete, widespread threat that leverages social engineering to acquire confidential data from victim users~\cite{Beardsley2005}. 
Phishing scams are often part of a profit-driven economy, where stolen data is sold in underground markets~\cite{Han2016,Bursztein2014}. They may be even used to achieve political or military objectives~\cite{Hong2012,Khonji2013}.  To maximize profit, as most of the current cybercrime activities, modern phishing attacks are  automatically deployed on a large scale, exploiting vulnerabilities in publicly-available websites through the so-called~\emph{phishing kits}~\cite{Han2016,Bursztein2014,Cova2008,Invernizzi2012,APWG2015}. These toolkits automatize the creation of phishing webpages on hijacked legitimate websites, and advertise the newly-created phishing sites to attract potential victims using dedicated spam campaigns.
The data harvested by the phishing campaign is then typically sold on the black market, and part of the profit is reinvested to further support the scam campaign~\cite{Han2016,Bursztein2014}.
To realize the importance of such a large-scale underground economy, note that, according to the most recent Global Phishing Survey by APWG, published in 2014, $59,485$ out of the $87,901$ domains linked to phishing scams (\ie, the $71.4\%$) were actually pointing to legitimate (compromised) websites~\cite{APWG2015}.

\begin{figure*}[t]
\centering
\includegraphics[height=0.25\textwidth]{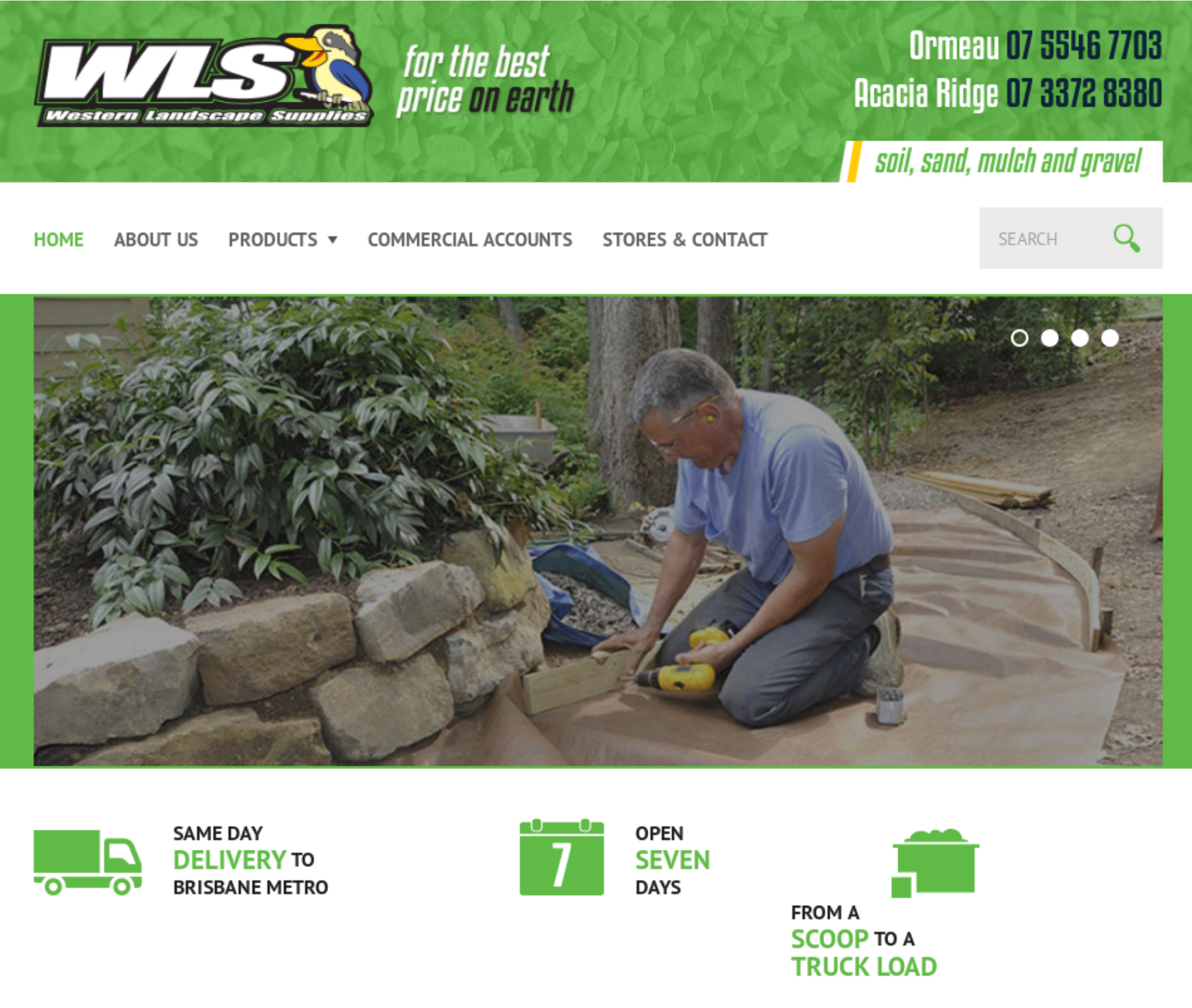} \hspace{1pt}
\includegraphics[height=0.25\textwidth]{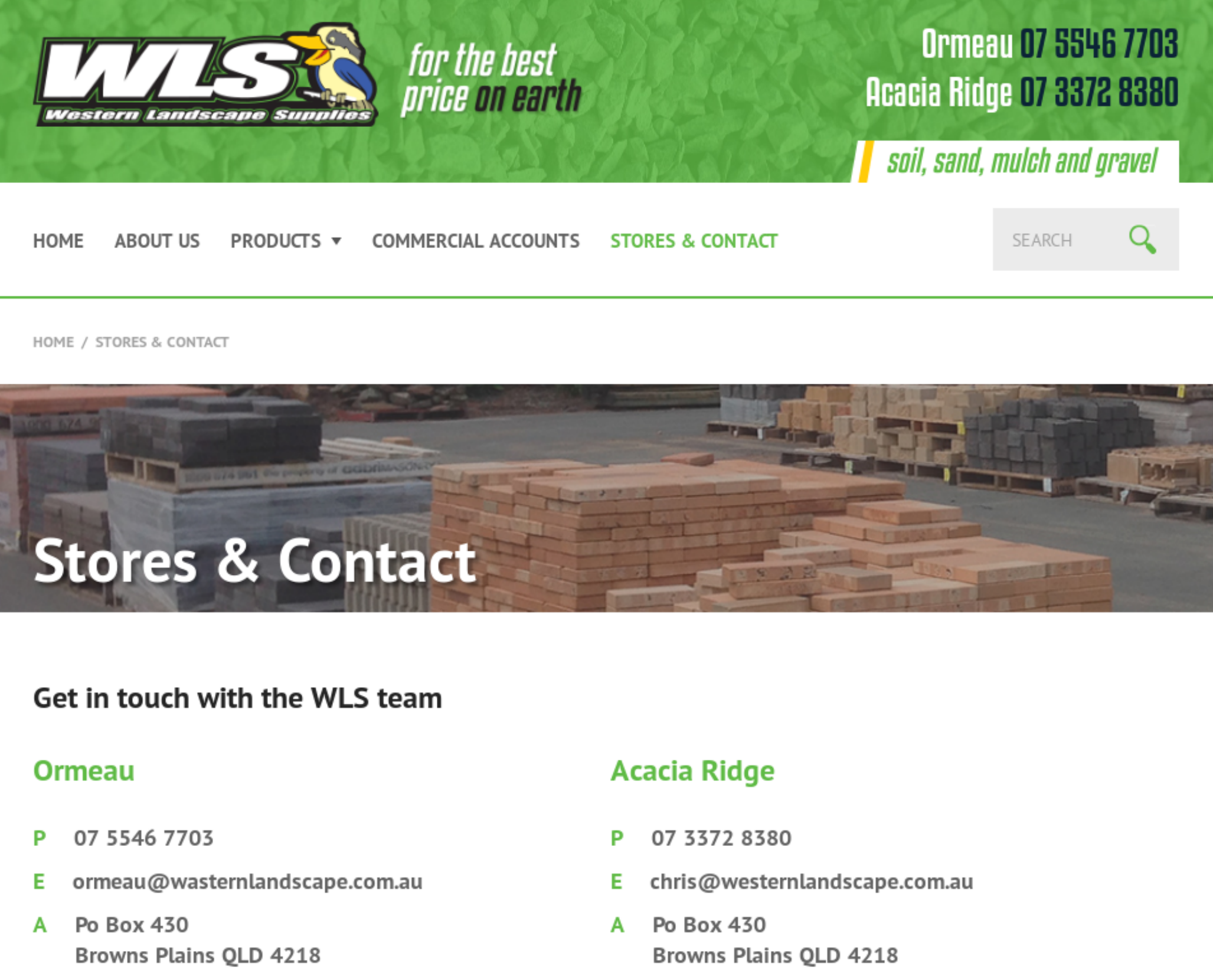} \hspace{1pt}
\includegraphics[height=0.25\textwidth]{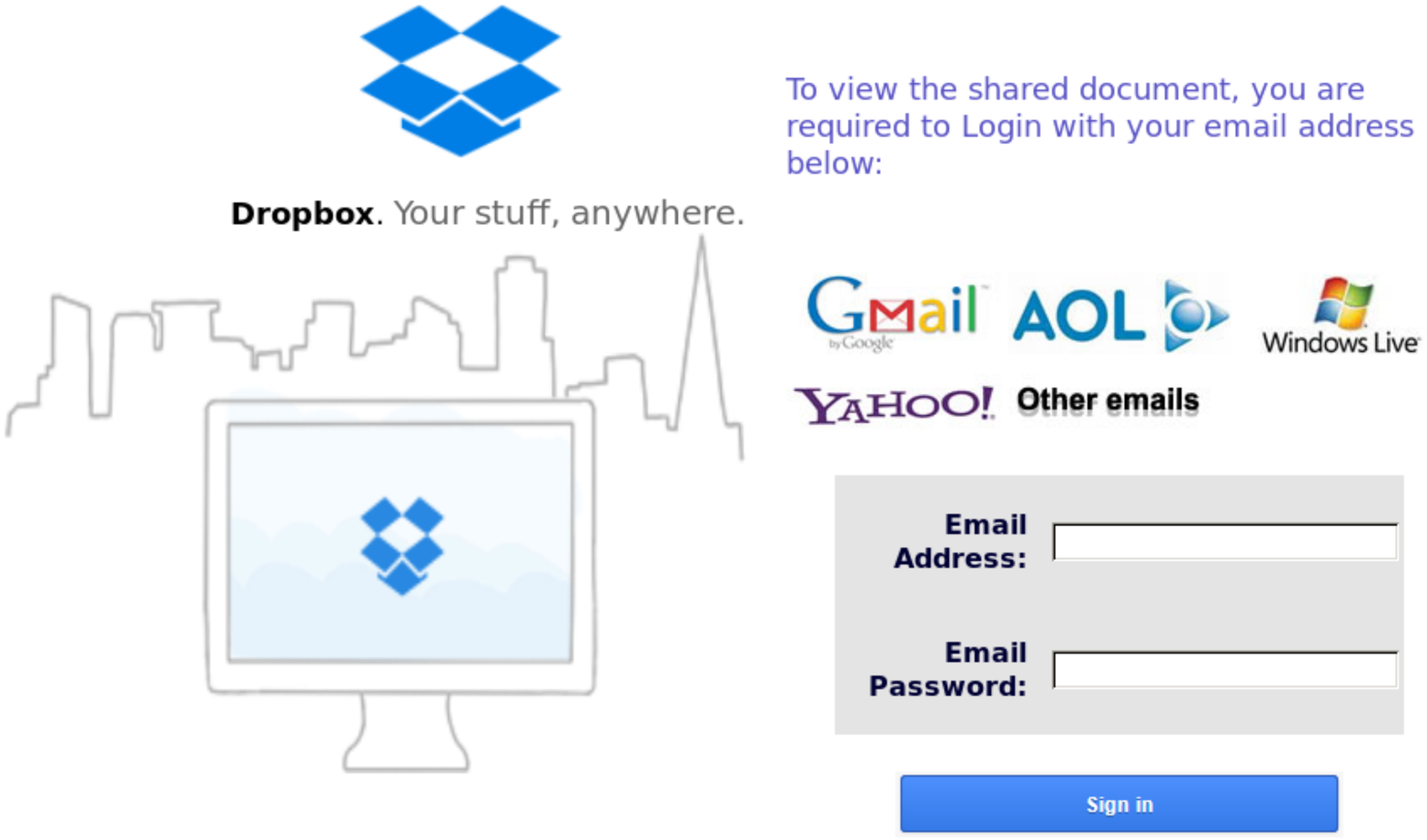}
\caption{Homepage (\emph{left}), legitimate (\emph{middle}) and phishing (\emph{right}) pages hosted in a compromised website.}
\label{fig:examples}
\end{figure*}

Compromising vulnerable, legitimate websites does not only enable a large-scale deployment of phishing attacks; it also provides several other advantages for cyber-criminals.
First, it does not require them to take care of registering domains and deal with hosting services to deploy their scam. This also circumvents recent approaches that detect malicious domains by evaluating abnormal domain behaviors (\eg, burst registrations, typosquatting domain names), induced by the need of automatizing domain registration~\cite{hao16-ccs}. 
On the other hand, website compromise is only a \emph{pivoting} step towards the final goal of the phishing scam. In fact, cyber-criminals normally leave the \emph{legitimate} pages hosted in the compromised website \emph{intact}. This allows them to hide the presence of website compromise not only from the eyes of its legitimate owner and users, but also from blacklisting mechanisms and browser plug-ins that rely on reputation services (as legitimate sites tend to have a good reputation)~\cite{Han2016}. 

For these reasons, malicious webpages in compromised websites remain typically undetected for a longer period of time. This has also been highlighted in a recent study by Han~\etal~\cite{Han2016}, in which the authors have exposed vulnerable websites (\ie, honeypots) to host and monitor phishing toolkits. 
They have reported that the first victims usually connect to phishing webpages within a couple of days after the hosting website has been compromised, while the phishing website is blacklisted by common services like \texttt{Google Safe Browsing} and \texttt{PhishTank} after approximately twelve days, on average. 
The same authors have also pointed out that the most sophisticated phishing kits include functionalities to evade blacklisting mechanisms. The idea is to redirect the victim to a randomly-generated subfolder within the compromised website, where the attacker has previously installed another copy of the phishing kit. 
Even if the victim realizes that he/she is visiting a phishing webpage, he/she will be likely to report the randomly-generated URL of the visited webpage (and not that of the redirecting one), which clearly makes blacklisting unable to stop this scam.

To date, several approaches have been proposed for phishing webpage detection (Sect.~\ref{sect:rel-work}). Most of them are based on comparing the candidate phishing webpage against a set of known targets~\cite{Basnet2014,Medvet2008}, 
or on extracting some generic features to discriminate between phishing and legitimate webpages~\cite{Chen2014,Blum2010}. 

To our knowledge, this is the first work that leverages the adversarial behavior of cyber-criminals to detect phishing pages in compromised websites, while overcoming some limitations of previous work. The key idea behind our approach, named \texttt{DeltaPhish} (or \deltaphish, for short), is to compare the HTML code and the \emph{visual} appearance of potential phishing pages against the corresponding characteristics of the homepage of the compromised (hosting) website (Sect.~\ref{sect:deltaphish}). In fact, phishing pages normally exhibit a much significant difference in terms of aspect and structure with respect to the website homepage than the other \emph{legitimate} pages of the website. The underlying reason is that phishing pages should resemble the appearance of the website targeted by the scam, while legitimate pages typically share the same style and aspect of their homepage (see, \eg, Fig.~\ref{fig:examples}).

Our approach is also robust to well-crafted manipulations of the HTML code of the phishing page, aimed to evade detection, as those performed in~\cite{Liang2016} to mislead the Google's Phishing Pages Filter embedded in the \emph{Chrome} web browser.
This is achieved by the proposal of two distinct \emph{adversarial fusion} schemes that combine the outputs of our HTML and visual analyses while accounting for potential attacks against them.
We consider attacks targeting the HTML code of the phishing page as altering also its visual appearance may significantly affect the effectiveness of the phishing scam. Preserving the visual similarity between a phishing page and the website targeted by the scam is indeed a fundamental \emph{trust-building} tactic used by miscreants to attract new victims~\cite{Beardsley2005}. 

In Sect.~\ref{sect:exp}, we simulate a case study in which \deltaphish is deployed as a module of a web application firewall, used to protect a specific website. In this setting, our approach can be used to detect whether users are accessing potential phishing webpages that are uploaded to the monitored website after its compromise. To simulate this scenario, we collect legitimate and phishing webpages hosted in compromised websites from \texttt{PhishTank}, and compare each of them with the corresponding homepage (which can be set as the reference page for \deltaphish when configuring the web application firewall).
We show that, under this setting, \deltaphish is able to correctly detect more than 99\% of the phishing pages while misclassifying less than 1\% of legitimate pages. We also show that \deltaphish can retain detection rates higher than $70\%$ even in the presence of adversarial attacks carefully crafted to evade it. To encourage reproducibility of our research, we have also made our dataset of $1,012$ phishing and $4,499$ legitimate webpages publicly available, along with the classification results of \deltaphish.

We conclude our work in Sect.~\ref{sect:conclusions}, highlighting its main limitations and related open issues for future research.

\section{Phishing Webpage Detection}
\label{sect:rel-work}

We categorize here previous work on the detection of phishing webpages along two main axes, depending on $(i)$ the detection approach, and $(ii)$ the features used for classification. The detection approach can be \emph{target-independent}, if it exploits generic features to discriminate between phishing and legitimate webpages, or \emph{target-dependent}, if it compares the suspect phishing webpage against known phishing targets. In both cases, features can be extracted from the webpage URL, its HTML content and visual appearance, as detailed below. 

\myparagraph{Target-independent.} These approaches exploit features computed from the webpage URL and its domain name~\cite{Garera2007,Blum2010,Le2011,Marchal2012}, 
from its HTML content and structure, 
and from other sources, including search engines, HTTP cookies, website certificates~\cite{Pan2006,Xu2013,Basnet2014,Whittaker2010,Xiang2010,Xiang2011,Britt2012,Jo2010}, 
 and even publicly-available blacklisting services like \texttt{Google Safe Browsing} and \texttt{PhishTank}~\cite{Ludl2007}.
Another line of work has considered the detection of phishing emails by analyzing their content along with that of the linked phishing webpages~\cite{Fette2007}.

\myparagraph{Target-dependent.} These techniques typically compare the potential phishing page to a set of known targets (\eg, \texttt{PayPal}, \texttt{eBay}).
HTML analysis has also been exploited to this end, often complemented by the use of search engines to identify phishing pages with similar text and page layout~\cite{Britt2012,Wardman2011}, 
or by the analysis of the pages linked to (or by) the suspect pages~\cite{Wenyin2012}. 
The main difference with target-independent approaches is that most of the target-dependent approaches have considered measures of \emph{visual similarity} between webpage \emph{snapshots} or embedded images, using a wide range of image analysis techniques, mostly based on computing low-level visual features, including color histograms, two-dimensional Haar wavelets, and other well-known image descriptors normally exploited in the field of computer vision~\cite{Chen2009a,Fu2006,Chen2014,Chen2010}. 
Notably, only few work has considered the combination of both HTML and visual characteristics~\cite{Medvet2008,Afroz2011}. 

\myparagraph{Limitations and Open Issues.}  The main limitations of current approaches and the related open research issues can be summarized as follows.
Despite \emph{target-dependent} approaches are normally more effective than \emph{target-independent} ones, they require a-priori knowledge of the set of websites that may be potentially targeted by phishing scams, or anyway try to retrieve them during operation by querying search engines.
This makes them clearly unable to detect phishing scams against unknown, legitimate services.
On the other hand, \emph{target-independent} techniques are, in principle, easier to evade, as they exploit generic characteristics of webpages to discriminate between phishing and legitimate pages, instead of making an explicit comparison between webpages. In particular, as shown in~\cite{Liang2016}, it is not only possible to infer enough information on how a publicly-available, \emph{target-independent} anti-phishing filter (like Google's Phishing Pages Filter) works, but it is also possible to exploit this information to evade detection, by carefully manipulating phishing webpages to resemble the characteristics of the legitimate webpages used to learn the classification system.
Evasion becomes clearly more difficult if visual analysis is also performed, as modifying the visual appearance of the phishing page tends to compromise the effectiveness of the phishing scam~\cite{Beardsley2005}.
However, mainly due to the higher computational complexity of this kind of analysis, only few approaches have combined HTML and visual features for target-dependent phishing detection~\cite{Medvet2008,Afroz2011}, 
and it is not clear to which extent they can be robust against well-crafted adversarial attacks.
Another relevant limitation is that no dataset has been made publicly available for comparing different detection approaches to a common benchmark, and this clearly hinders research reproducibility.

Our approach overcomes many of the aforementioned limitations. First, it does not require any knowledge of legitimate websites potentially targeted by phishing scams.
Although it may be thus considered a target-independent approach, it is not based on extracting generic features from phishing and legitimate webpages, but rather on comparing the characteristics of the phishing page to those of the homepage hosted in the compromised website. 
This makes it more robust than other target-independent approaches against evasion attempts in which, \eg, the HTML code of the phishing webpage is obfuscated, as this would make the phishing webpage even more \emph{different} from the homepage.
Furthermore, we explicitly consider a security-by-design approach while engineering our system, based on explicitly accounting for well-crafted attacks against it. As we will show, our \emph{adversarial fusion} mechanisms guarantee high detection rates even under worst-case changes in the HTML code of phishing pages, by effectively leveraging the role of the visual analysis.
Finally, we publicly release our dataset to encourage research reproducibility and benchmarking.

\vspace{-10pt}
\section{DeltaPhish} \label{sect:deltaphish}
\vspace{-5pt}

\begin{figure*}[t]
\begin{center}
\includegraphics[width=0.9\textwidth]{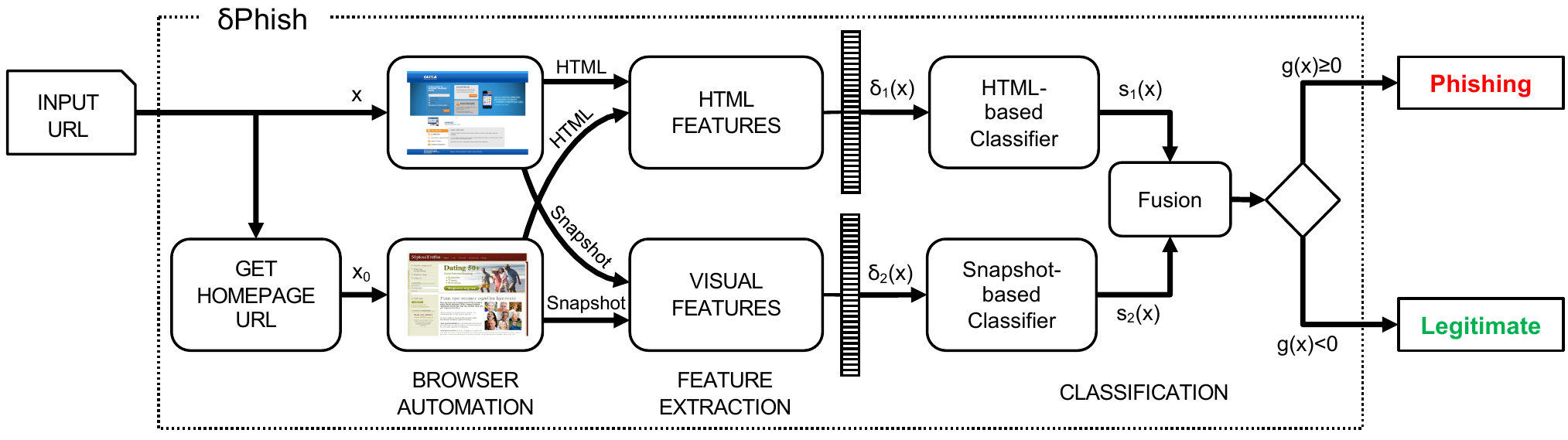}
\caption{High-level architecture of \deltaphish.}
\label{fig:detection}
\end{center}
\end{figure*}

In this section we present \texttt{DeltaPhish} (\deltaphish). Its name derives from the fact that it determines whether a certain URL contains a phishing webpage by evaluating HTML and visual \emph{differences} between the input page and the website homepage.
The general architecture of \deltaphish is depicted in Fig.~\ref{fig:detection}.
We denote with $x \in \set X$ either the URL of the input webpage or the webpage itself, interchangeably. 
Accordingly, the set $\set X$ represents all possible URLs or webpages. The homepage hosted in the same domain of the visited page (or its URL) is denoted with $x_{0} \in \set X$.
Initially, our system receives the input URL of the input webpage $x$ and retrieves that of the corresponding homepage $x_{0}$.
Each of these URLs is received as input by a \emph{browser automation} module (Sect.~\ref{sect:bro-auto}), which downloads the corresponding page and outputs its HTML code and a snapshot image.
The HTML code of the input page and that of the homepage are then used to compute a set of HTML features (Sect.~\ref{sub-sec:HTML-Based}).
Similarly, the two snapshot images are passed to another feature extractor that computes a set of visual features (Sect.~\ref{sub-sec:Snapshot-Based}).
The goal of these feature extractors is to map the input page $x$ onto a vector space suitable for learning a classification function.
Recall that both feature sets are computed based on a \emph{comparison} between the characteristics of the input page $x$ and those of the homepage $x_{0}$.
We denote the two mapping functions implemented by the HTML and by the visual feature extractor respectively with $\delta_{1}(x) \in \mathbb R^{\con d_{1}}$ and $\delta_{2}(x) \in \mathbb R^{\con d_{2}}$, being $\con d_{1}, \con d_{2}$ the dimensionality of the two vector spaces.
For compactness of our notation, we do not explicitly highlight the dependency of $\delta_{1}(x)$ and $\delta_{2}(x)$ on $x_{0}$, even if it should be clear that such functions depend on both $x$ and $x_{0}$.
These two vectorial-based representations are then used to learn two distinct classifiers, \ie, an HTML- and a Snapshot-based classifier. During operation, these classifiers will respectively output a \emph{dissimilarity} score $s_{1}(x) \in \mathbb R$ and $s_{2}(x) \in \mathbb R$ for each input page $x$, which essentially measure how \emph{different} the input page is from the corresponding homepage. Thus, the higher the score, the higher the probability of $x$ being a phishing page. These scores are then combined using different (standard and adversarial) \emph{fusion} schemes (Sect.~\ref{sub-sec:class-fusion}), to output an aggregated score $g(x) \in \mathbb R$. If $g(x) \geq 0$, the input page $x$ is classified as a phish, and as legitimate otherwise.  

Before delving into the technical implementation of each module, it is worth remarking that \deltaphish can be implemented as a module in web application firewalls, and, potentially, also as an online blacklisting service (to filter suspicious URLs). Some implementation details that can be used to speed up the processing time of our approach are discussed in Sect.~\ref{sect:exp-res}.

\subsection{Browser Automation}
\label{sect:bro-auto}

The browser automation module launches a browser instance using \emph{Selenium}\footnote{\url{http://docs.seleniumhq.org}} to gather the snapshot of the landing web page and its HTML source, even if the latter is dynamically generated with (obfuscated) JavaScript code. This is indeed a common case for phishing webpages.

\subsection{HTML-Based Classification}
\label{sub-sec:HTML-Based}

For HTML-based classification, we define a set of $11$ features, obtained by comparing the input page $x$ and the homepage $x_{0}$ of the website hosted in the same domain. They will be the elements of the $\con d_{1}$-dimensional feature vector $\delta_{1}(x)$ (with $\con d_{1}=11$) depicted in Fig.~\ref{fig:detection}.
We use the Jaccard index $J$ as a similarity measure to compute most of the feature values. Given two sets $A, B$, it is  defined as the cardinality of their intersection divided by the cardinality of their union:
\begin{eqnarray}\vspace{-5pt}
	J(A, B)={\lvert A \cap B \lvert} / {\lvert A \cup B \lvert} \in [0,1] \, .
\label{eq:jaccard}\vspace{-5pt}
\end{eqnarray}
If $A$ and $B$ are both empty, $J(A,B)=1$.
The $11$ HTML features used by our approach are described below.

\myparagraph{$(1)$ URL.} We extract all URLs corresponding to hyperlinks in $x$ and $x_{0}$ through the inspection of the \texttt{href} attribute of the \texttt{<a>} tag,\footnote{Recall that the \texttt{<a>} tag defines a hyperlink and the \texttt{href} attribute is its destination.} and create a set for each page. URLs are considered once in each set without repetition. 
We then compute the Jaccard index (Eq.~\ref{eq:jaccard}) of the two sets extracted. 
For instance, let us assume that $x$ and $x_{0}$ respectively contain these two URL sets:
\begin{enumerate}
	\item[$U_{x}:$] \begin{itemize}
		\item[] \{\texttt{https://www.example.com/p1/}, \texttt{https://www.example.com/p2/},
		\item[] \texttt{https://support.example.com/}\}
	\end{itemize}
	\item[$U_{x_{0}}:$] \begin{itemize}
		\item[] \{\texttt{https://support.example.com/p1}, \texttt{https://www.example.com/p2/},
		\item[] \texttt{https://support.example.com/en-us/ht20}\}
	\end{itemize}
\end{enumerate}
In this case, since only one element is exactly the same in both sets (\ie, \texttt{https://www.example.com/p2/}), the Jaccard index is $J(U_{x}, U_{x_{0}})=0.2$.

\myparagraph{$(2)$ 2LD.} This feature is similar to the previous one, except that we consider the second-level domains (2LDs) extracted from each URL instead of the full link. The 2LDs are considered once in each set without repetition. Let us now consider the example given for the computation of the previous feature. In this case, both $U_{x}$ and $U_{x_{0}}$ will contain only \texttt{example.com}, and, thus, $J(U_{x}, U_{x_{0}})=1$.

\myparagraph{$(3)$ SS.} To compute this feature, we extract the content of the \texttt{<style>} tags from $x$ and $x_{0}$. They are used to define style information, and every webpage can embed multiple \texttt{<style>} tags. We compare the similarity between the sets of \texttt{<style>} tags of $x$ and $x_{0}$ using the Jaccard index. 
%

\myparagraph{$(4)$ SS-URL.}  We extract URLs from $x$ and $x_{0}$ that point to external style sheets through the inspection of the \texttt{href} attribute of the \texttt{<link>} tag; 
\eg, \texttt{http://example.com/resources/styles.css}.
We create a set of URLs for $x$ and another for $x_{0}$ (where every URL appears once in each set, without repetition), and compute their similarity using the Jaccard index (Eq.~\ref{eq:jaccard}).

\myparagraph{$(5)$ SS-2LD.} As for the previous feature, we extract all the URLs that link external style sheets in $x$ and $x_{0}$. However, in this case we only consider the second-level domains for each URL (\eg, \texttt{example.com}). The feature value is then computed again using the Jaccard index (Eq.~\ref{eq:jaccard}).

\myparagraph{$(6)$ I-URL.} For this feature, we consider the URLs of linked images in $x$ and in $x_{0}$, separately, by extracting all the URLs  specified in the \texttt{<img src=\ldots>} attributes. The elements of these two sets are image URLs; \\ 
\eg, \texttt{http://example.com/img/image.jpg}, and are considered once in each set without repetition. We then compute the Jaccard index for these two sets (Eq.~\ref{eq:jaccard}).

\myparagraph{$(7)$ I-2LD.} We consider the same image URLs extracted for \textbf{I-URL}, but restricted to their 2LDs. Each 2LD is considered once in each set without repetition, and the feature value is computed using again the Jaccard index (Eq.~\ref{eq:jaccard}).

\myparagraph{$(8)$ Copyright.} We extract all significant words, sentences and symbols found in $x$ and $x_{0}$ that can be related to copyright claims (\eg, \textcopyright, \emph{copyright}, \emph{all rights reserved}), without  repetitions, and excluding stop-words of all human languages. The feature value is then computed using the Jaccard index.

\myparagraph{$(9)$ X-links.} This is a binary feature. It equals $1$ if the homepage $x_{0}$ is linked in $x$ (accounting for potential redirections), and $0$ otherwise. 

\myparagraph{$(10)$ Title.} This feature is also computed using the Jaccard index. We create the two sets to be compared by extracting all words (except stop-words) from the title of $x$ and $x_{0}$, respectively, without repetitions.
They can be found within the tag \texttt{<title>}, which defines the title of the HTML document, \ie, the one appearing on the browser toolbar and displayed in search-engine results. 

\myparagraph{$(11)$ Language.} This feature is set to $1$ if $x$ and $x_{0}$ use the same language, and to $0$ otherwise. To identify the language of a page, we first extract the stop-words for all the human languages known from $x$ and $x_{0}$, separately, and without repetitions. We then assume that the page language is that associated to the maximum number of corresponding stop-words found.

\myparagraph{Classification.} The $11$ HTML features map our input page $x$ onto a vector space suitable for classification. Using the compact notation defined at the beginning of this section (see also Fig.~\ref{fig:detection}), we denote the  $\con d_{1}$-dimensional feature vector corresponding to $x$ as $\delta_{1}(x)$ (being $\con d_{1}=11$).
We then train a linear Support Vector Machine (SVM)~\cite{vapnik95} on these features to classify phishing and legitimate pages. For each input page, during operation, this classifier computes a \emph{dissimilarity score} measuring how different the input page is from its homepage:
\begin{equation}
s_{1}(x) = \vct w_{1}^{T} \delta_{1}(x) + b_{1} \, .
\label{eq:s1}
\end{equation}
The feature weights $\vct w_{1} \in \mathbb R^{\con d_{1}}$ and the bias $b_{1} \in \mathbb R$ of the classification function are optimized during SVM learning, using a labeled set of training webpages~\cite{vapnik95}.

\subsection{Snapshot-Based Classification}
\label{sub-sec:Snapshot-Based}

\begin{figure*}[t]
\centering
\includegraphics[width=0.9\textwidth]{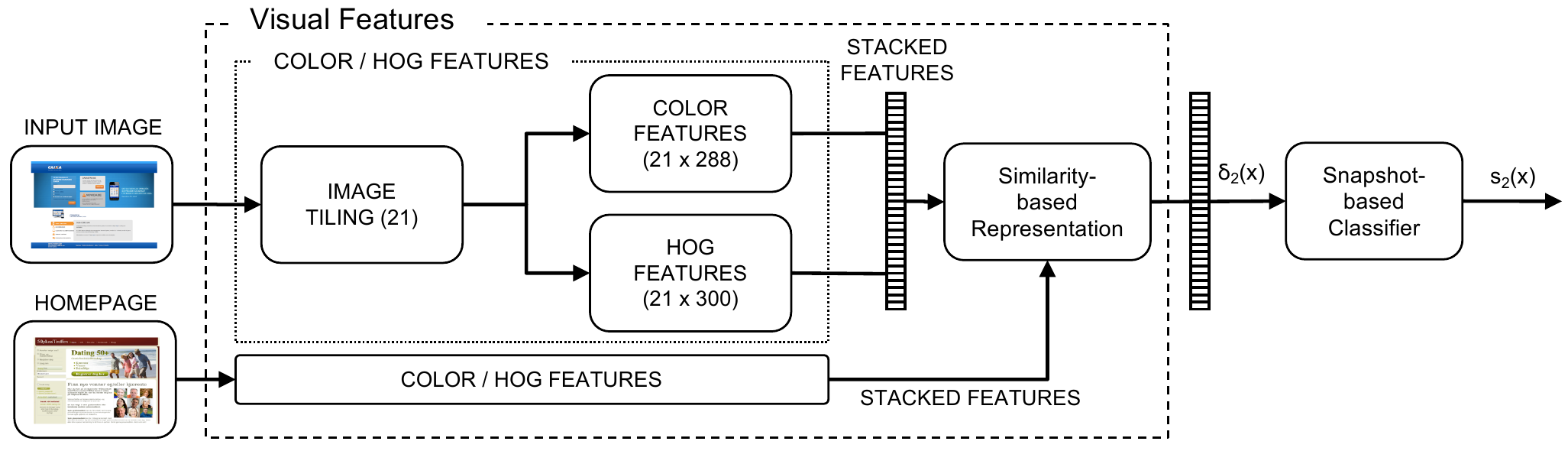}
\caption{Computation of the \emph{visual features} in \deltaphish.}
\label{fig:visual-feat}
\end{figure*}
To analyze differences in the snapshots of the input page $x$ and the corresponding homepage $x_{0}$,
we leverage two state-of-the-art feature representations that are widely used for image classification, \ie, the so-called Histogram of Oriented Gradients (HOGs)~\cite{DalalT05}, and color histograms. 
We have selected these features since, with respect to other popular descriptors (like the Scale-Invariant Feature Transform, SIFT), they typically achieve better performance in the presence of very high inter-class similarities.
Unlike HOGs, which are local descriptors, color histograms give a representation of the spatial distribution of colors within an image, providing complementary information to our snapshot analysis.

We exploit these two representations to compute a concatenated (stacked) feature vector for each snapshot image, and then define a way to compute a similarity-based representation from them. The overall architecture of our snapshot-based classifier is depicted in Fig.~\ref{fig:visual-feat}.
In the following, we explain more in detail how HOG and color histograms are computed for each snapshot image separately, and how we combine the stacked feature vectors of the input page $x$ and of the homepage $x_{0}$ to obtain the final similarity-based feature vector.

\begin{figure*}[t]
\centering
\includegraphics[width=0.85\textwidth]{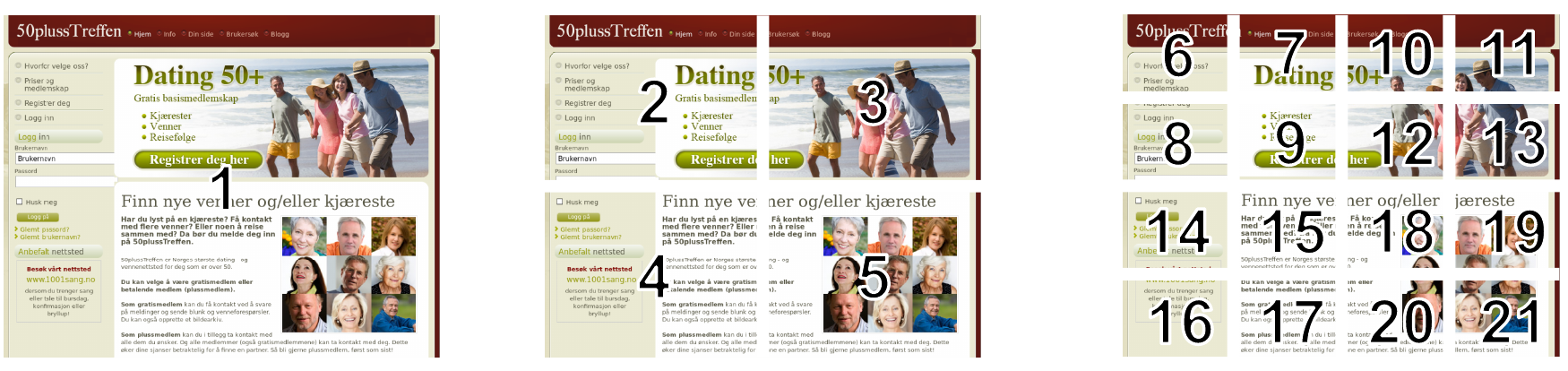}
\caption{\deltaphish image tiling extracts visual features retaining spatial information.}
\label{fig:Image_tiling}
\end{figure*}

\myparagraph{Image Tiling.} To preserve spatial information in our visual representation of the snapshot, we extract visual features not only from the whole snapshot image, but also from its quarters and sixteenths (as depicted in Fig.~\ref{fig:Image_tiling}), yielding $(1 \times 1)+(2 \times 2)+(4 \times 4) = 21$ tiles. HOG descriptors and color histograms are extracted from each tile, and stacked, to obtain two vectors of $21 \times 300=6,300$ and $21 \times 288=6,048$ dimensions, respectively. 

\begin{figure*}[t]
\centering
\includegraphics[width=0.9\textwidth]{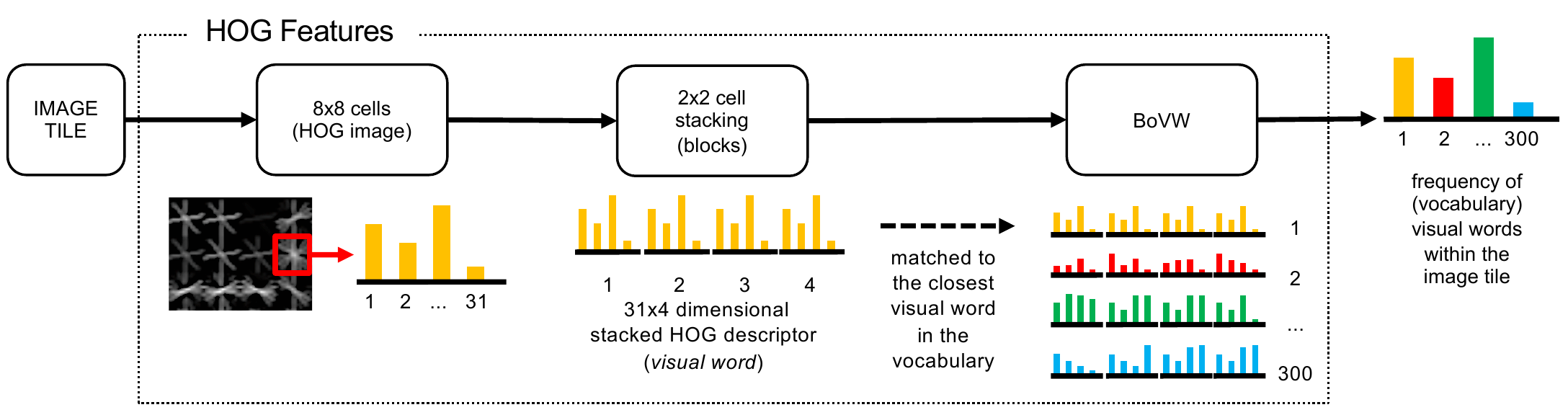}
\caption{Computation of the $300$ HOG features from an image tile. 
}
\label{fig:HOG_extraction}
\end{figure*}

\myparagraph{HOG features.} We compute the HOG features for each of the $21$ input \emph{image tiles} following the steps highlighted in Fig.~\ref{fig:HOG_extraction} and detailed below, as in~\cite{DalalT05}.
First, the image is divided in cells of $8\times8$ pixels.
For each cell, a 31-dimensional HOG descriptor is computed, in which each bin  represents a quantized direction and its value corresponds to the magnitude of gradients in that direction (we refer the reader to~\cite{VedaldiF10,DalalT05,FelzenszwalbGMR10} for further details).
%
The second step consists of considering overlapping blocks of $2\times2$ neighboring cells (\ie, $16 \times 16$ pixels).
For each block, the 31-dimensional HOG descriptors of the four cells are simply concatenated to form a ($31\times 4$) 124-dimensional \emph{stacked} HOG descriptor, also referred to as a \emph{visual word}.
In the third step, each visual word extracted from the image tile is compared against a pre-computed vocabulary of $K$ visual words, and assigned to the closest word in the vocabulary (we have used $K=300$ visual words in our experiments). Eventually, a histogram of $K=300$ bins is obtained for the whole tile image, where each bin represents the occurrence of each pre-computed visual word in the tile. This approach is usually referred to as Bag of Visual Words (BoVW)~\cite{Csurka04visualcategorization}. The vocabulary can be built using the centroids found by $k$-means clustering from the whole set of visual words in the training data. Alternatively, a vocabulary computed from a different dataset may be also used.

\myparagraph{Color features.} To extract our color features, we first convert the image from the RGB (Red-Green-Blue) to the HSV (Hue-Saturation-Value) color space, and perform the same image tiling done for the extraction of the HOG features (see Fig.~\ref{fig:Image_tiling}).
We then compute a quantized 3D color histogram with 8, 12 and 3 bins respectively for the H, S and V channel, corresponding to a vector of $8 \times 12 \times 3 = 288$ feature values. This technique has shown to be capable of outperforming histograms computed in the RGB color space, in content-based image retrieval and image segmentation tasks~\cite{SuralQP02a}.

Both the HOG descriptor and the color histogram obtained from each image tile are normalized to sum up to one (to correctly represent the relative frequency of each bin). The resulting $21 \times 300$ HOG descriptors and $21 \times 288$ color histograms are then stacked to obtain a feature vector consisting of $\con d_{2} = 12,348$ feature values, as shown in Fig.~\ref{fig:visual-feat}.
In the following, we denote this vector respectively with $\vct p$ and $\vct p_{0}$ for the input page $x$ and the homepage $x_{0}$.

\myparagraph{Similarity-based Feature Representation.} After computing the visual features $\vct p$ for the input page $x$ and $\vct p_{0}$ for the homepage $x_{0}$, we compute the similarity-based representation $\delta_{2}(x)$ (Figs.~\ref{fig:detection}-\ref{fig:visual-feat}) from these feature vectors as:
\begin{equation}
\delta_{2}(x) = \min (\vct p , \vct p_{0}) \,
\label{eq:d2}\vspace{-2.5pt}
\end{equation}
where $\min$ here returns the minimum of the two vectors for each coordinate. Thus, the vector $\delta_{2}(x)$ will also consists of $\con d_{2} = 12,348$ feature values. 

\myparagraph{Classification.}  The similarity-based mapping in Eq.~\eqref{eq:d2} is inspired to the histogram intersection kernel~\cite{SwainB91}. This kernel evaluates the similarity between two histograms $u$ and $v$ as $\sum_{i} \min (u_{i}, v_{i})$. Instead of summing up the values of $\delta_{2}(x)$ (which will give us exactly the histogram intersection kernel between the input page and the homepage), we learn a linear SVM to estimate a weighted sum:
\begin{equation}
s_{2}(x) = \vct w_{2}^{T} \delta_{2}(x) + b_{2} \, ,
\label{eq:s2}\vspace{-2.5pt}
\end{equation}
where, similarly to the HTML-based classifier, $\vct w_{2} \in \mathbb R^{\con d_{2}}$ and $b_{2} \in \mathbb R$ are the feature weights and bias, respectively.
This enables us to achieve better performances, as, in practice, the classifier itself  learns a proper similarity measure between webpages directly from the training data. This is a well-known practice in the area of machine learning, usually referred to as \emph{similarity learning}~\cite{chechik10}.

\vspace{-5pt}
\subsection{Classifier Fusion}\label{sub-sec:class-fusion}
\vspace{-2pt}

The outputs of the HTML- and of the Snapshot-based classifiers, denoted in the following with a two-dimensional vector $\vct s = (s_{1}(x), s_{2}(x))$ (Eqs.~\ref{eq:s1}-\ref{eq:s2}), can be combined using a fixed (untrained) fusion rule, or a classifier (trained fusion).
We consider three different combiners in our experiments, as described below.

\myparagraph{Maximum.} This rule simply computes the overall score as: 
\begin{equation}
g(x) = \max \left(s_{1}(x) ,s_{2}(x) \right ) \, . 
\label{eq:max}\vspace{-2pt}
\end{equation}
The idea is that, for a page to be classified as legitimate, both classifiers should output a low score. If one of the two classifiers outputs a high score and classifies the page as a phish, then the overall system will also classify it as a phishing page. The reason behind this choice relies upon the fact that the HTML-based classifier can be evaded by a skilled attacker, as we will see in our experiments, and we aim to avoid that misleading such a classifier will suffice to evade the whole system. In other words, we would like our system to be evaded only if both classifiers are successfully fooled by the attacker. For this reason, this simple rule can be also considered itself a sort of \emph{adversarial fusion} scheme.

\myparagraph{Trained Fusion.} To implement this fusion mechanism, we use an SVM with the Radial Basis Function (RBF) kernel, which computes the overall score as:
\begin{equation}
g(x) = \textstyle \sum_{i=1}^{\con n} y_{i} \alpha_{i} k( \vct s , \vct s_{i} ) + b\, , 
\label{eq:svm-rbf}\vspace{-2pt}
\end{equation}
where $k (\vct s, \vct s_{i}) = \exp{(-\gamma \| \vct s - \vct s_{i}\|^{2})}$ is the RBF kernel function, $\gamma$ is the kernel parameter, and $\vct s = (s_{1}(x), s_{2}(x))$ and $\vct s_{i} = (s_{1}(x_{i}), s_{2}(x_{i}))$ are the scores provided by the HTML- and Snapshot-based classifiers for the input page $x$ and for the $\con n$ pages in our training set $\set D = \{ x_{i}, y_{i}\}_{i=1}^{\con n}$, being $y_{i} \in \{-1,+1\}$ the class label (\ie, $-1$ and $+1$ for legitimate and phishing pages). The classifier parameters $\{\alpha_{i}\}_{i=1}^{\con n}$ and $b$ are estimated during training by the SVM learning algorithm, on the set of scores $\set S = \{\vct s_{i}, y_{i}\}_{i=1}^{\con n}$, which can be computed through \emph{stacked generalization} (to avoid overfitting~\cite{wolpert92}) as explained in Sect.~\ref{sect:exp-setup}.

\myparagraph{Adversarial Fusion.} In this case, we consider the same trained fusion mechanism described above, but augment the training scores by simulating attacks against the HTML-based classifier.
In particular, we add a fraction of samples for which the score of the Snapshot-based classifier is not altered, while the score of the HTML-based classifier is randomly sampled from a uniform distribution in $[0,1]$. This is a straightforward way to account for the fact that the score of the HTML-based classifier can be potentially decreased by a targeted attack against that module, and make the combiner aware of this potential threat.

Some examples of the resulting decision functions are shown in Fig.~\ref{fig:secure-fusion}. Worth remarking, when using trained fusion rules, the output scores of the the HTML- and Snapshot-based classifiers are normalized in $[0,1]$ using min-max normalization, to facilitate learning (see Sect.~\ref{sect:exp-setup} for further details).

\vspace{-8pt}
\section{Experimental Evaluation} \label{sect:exp}
\vspace{-5pt}
In this section we empirically evaluate \deltaphish, simulating its application as a module in a web application firewall. Under this scenario, we assume that the monitored website has been compromised (\eg, using a phishing kit), and it is hosting a phishing webpage.
The URLs contacted by users visiting the website are monitored by the web application firewall, which can deny access to a resource if retained suspicious (or which can stop a request if retained a potential attack against the web server). The contacted URLs that are not blocked by the web application firewall 
 are forwarded to \deltaphish, which detects whether they are substantially different from the homepage (\ie, they are potential phishing pages hosted in the monitored website). If \deltaphish reveals such a sign of compromise, the web application firewall can deny user access to the corresponding URL.

We first discuss the characteristics of the  webpages we have collected from legitimate, compromised websites (hosting phishing scams) to build our dataset, along with the settings used to run our experiments (Sect.~\ref{sect:exp-setup}). We then report our results, showing that our system can detect most of the phishing pages with very high accuracy, while misclassifying only few legitimate webpages (Sect.~\ref{sect:exp-res}). We have also considered an adversarial evaluation of our system in which the characteristics of the phishing pages are manipulated to evade detection of the HTML-based classifier. The goal of this adversarial analysis is to show that \deltaphish can successfully resist even to worst-case evasive attempts. Notably, we have not considered attacks against the Snapshot-based classifier as they would require modifying the visual aspect of the phishing page, thus making it easier for the victim to recognize the phishing scam.

\vspace{-5pt}
\subsection{Experimental Setting} \label{sect:exp-setup}
\vspace{-5pt}

\myparagraph{Dataset.} Our dataset has been collected from October 2015 to January 2016, starting from \emph{active} phishing URLs obtained online from the PhishTank feed.\footnote{\url{https://www.phishtank.com}}
We have collected and manually validated $1,012$ phishing pages.
For each phishing page, we have then collected the corresponding homepage from the hosting domain.
By parsing the hyperlinks in the HTML code of the homepage, we have collected from $3$ to $5$ legitimate pages from the same website, and validated them manually. This has allowed us to gather $1,012$ distinct sets of webpages, from now on referred to as \emph{families}, each consisting of a phishing page and some legitimate pages collected from the \emph{same} website.
Overall, our dataset consists of $5,511$ distinct webpages, $1,012$ of which are phishing pages.
We make this data publicly available, along with the classification results of \deltaphish.\footnote{\url{http://deltaphish.pluribus-one.it/}}

In these experiments, we consider 20 distinct training-test pairs to average our results. 
For a fair evaluation, webpages collected from the same domain (\ie, belonging to the same \emph{family}) are included either in the training data or in the test data.
In each repetition, we randomly select 60\% of the families for training, while the remaining 40\% are used for testing.
We normalize the feature values $\delta_{1}(x)$ and $\delta_{2}(x)$ using min-max normalization, but estimating the $5^{\rm th}$ and the $95^{\rm th}$ percentile from the training data for each feature value, instead of the minimum and the maximum, to reduce the influence of outlying feature values.

This setting corresponds to the case in which \deltaphish is trained before deployment on the web application firewall, to detect phishing webpages independently from the specific website being monitored. It is nevertheless worth pointing out that our system can also be trained using only the legitimate pages of the monitored website, \ie, it can be customized depending on the specific deployment.

\myparagraph{Classifiers.} We consider the HTML- and Snapshot-based classifiers (Sects.~\ref{sub-sec:HTML-Based}-\ref{sub-sec:Snapshot-Based}), using the three fusion rules discussed in Sect.~\ref{sub-sec:class-fusion} to combine their outputs:
$(i)$ \textbf{Fusion (max.)}, in which the $\max$ rule is used to combine the two outputs (Eq.~\ref{eq:max}); $(ii)$ \textbf{Fusion (tr.)}, in which we use an SVM with the RBF kernel as the combiner (Eq.~\ref{eq:svm-rbf}); and $(iii)$ \textbf{Fusion (adv.)}, in which we also use an SVM with the RBF kernel as the combiner, but augment the training set with phishing webpages \emph{adversarially manipulated} to evade the HTML-based classifier.

\myparagraph{Parameter tuning.} For HTML- and Snapshot-based classifiers, the only parameter to be tuned is the regularization parameter $C$ of the SVM algorithm. For SVM-based combiners exploiting the RBF kernel, we also have to set the kernel parameter $\gamma$.
In both cases, we exploit a 5-fold cross-validation procedure to tune the parameters, by performing a grid search on $C, \gamma \in \{0.001, 0.01, 0.1, 1, 10, 100\}$.
As the trained fusion rules require a separate training set for the base classifiers and the combiner (to avoid overfitting), we run a two-level (nested) cross-validation procedure, usually referred to as \emph{stacked generalization}~\cite{wolpert92}. In particular, the outer 5-fold cross validation splits the training data into a further training and validation set. This training set is used to tune the parameters (using an inner 5-fold cross validation as described above) and train the base classifiers. Then, these classifiers are evaluated on the validation data, and their outputs on each validation sample are stored.
We normalize these output scores in $[0,1]$ using min-max normalization.
At the end of the outer cross-validation procedure, 
we have computed the outputs of the base classifiers for each of the initial training samples, \ie, the set $\set S = \{\vct s_{i}, y_{i}\}_{i=1}^{\con n}$ (Sect.~\ref{sub-sec:class-fusion}).
We can thus optimize the parameters of the combiner on this data and then learn the fusion rule on all data.
For the adversarial fusion, we set the fraction of simulated attacks added to the training score set to 30\% (Sect.~\ref{sub-sec:class-fusion}).

\begin{figure*}[t]
\centering
\includegraphics[width=0.44\textwidth]{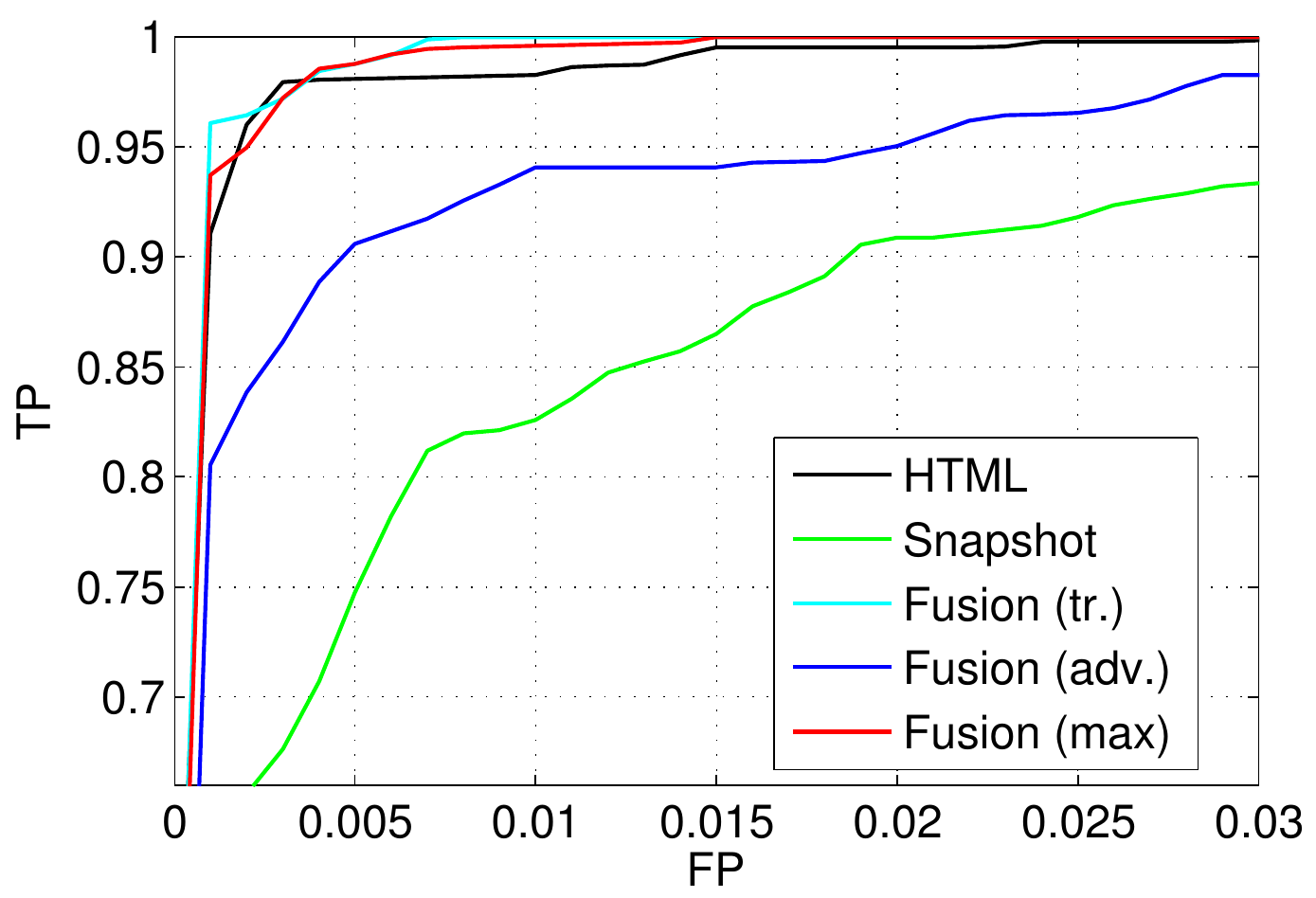}
\includegraphics[width=0.42\textwidth]{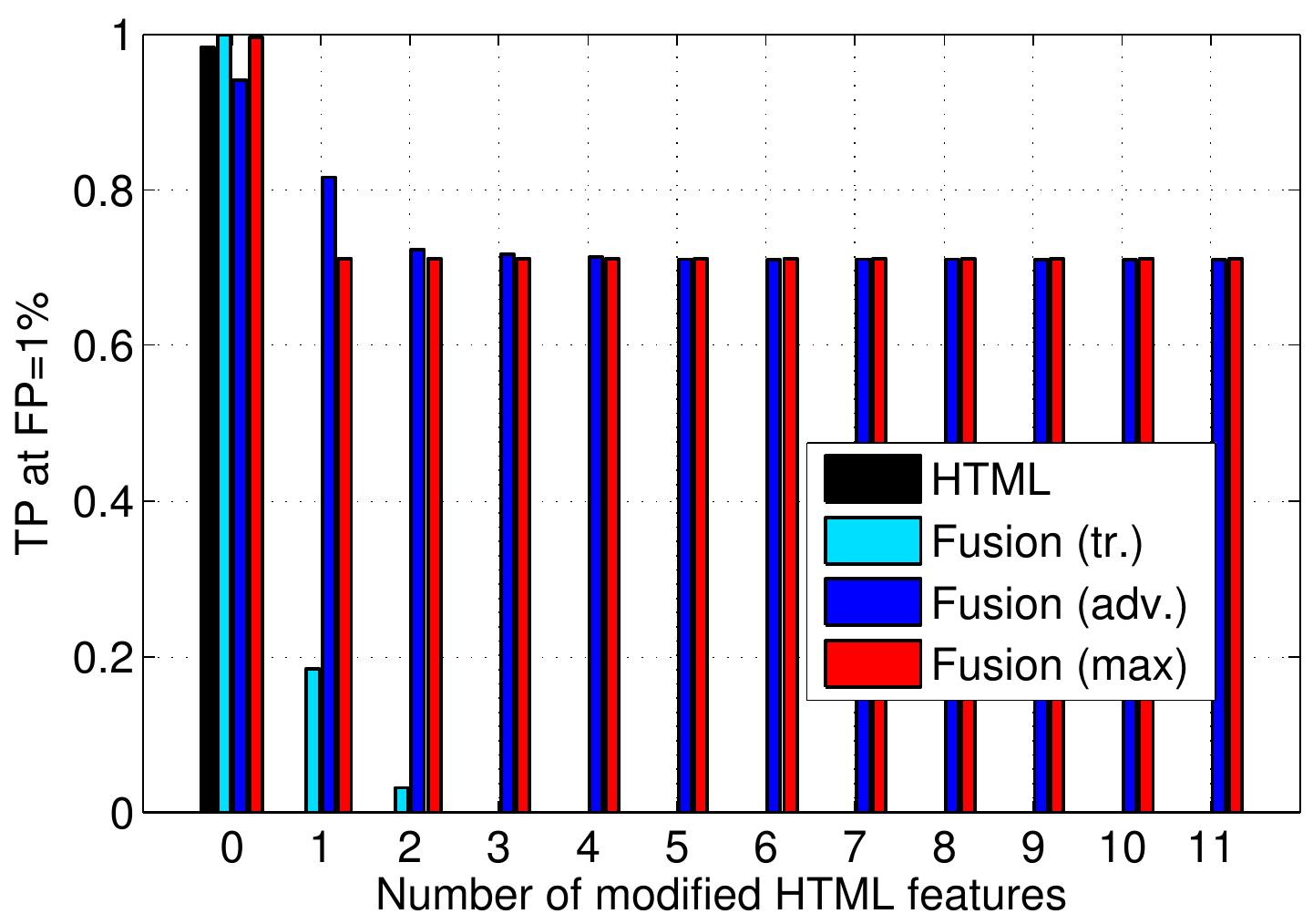}
\caption{ROC curves (\emph{left}) and adversarial evaluation (\emph{right}) of the classifiers.}
\label{fig:phish-det}
\end{figure*}

\subsection{Experimental Results} \label{sect:exp-res}

The results for phishing detection are shown in Fig.~\ref{fig:phish-det} (\emph{left} plot), using Receiver-Operating-Characteristic (ROC) curves. Each curve reports the average detection rate of phishing pages (\ie, the true positive rate, TP) against the fraction of misclassified legitimate pages (\ie, the false positive rate, FP).

\begin{figure*}[t]
\centering
\includegraphics[width=0.325\textwidth]{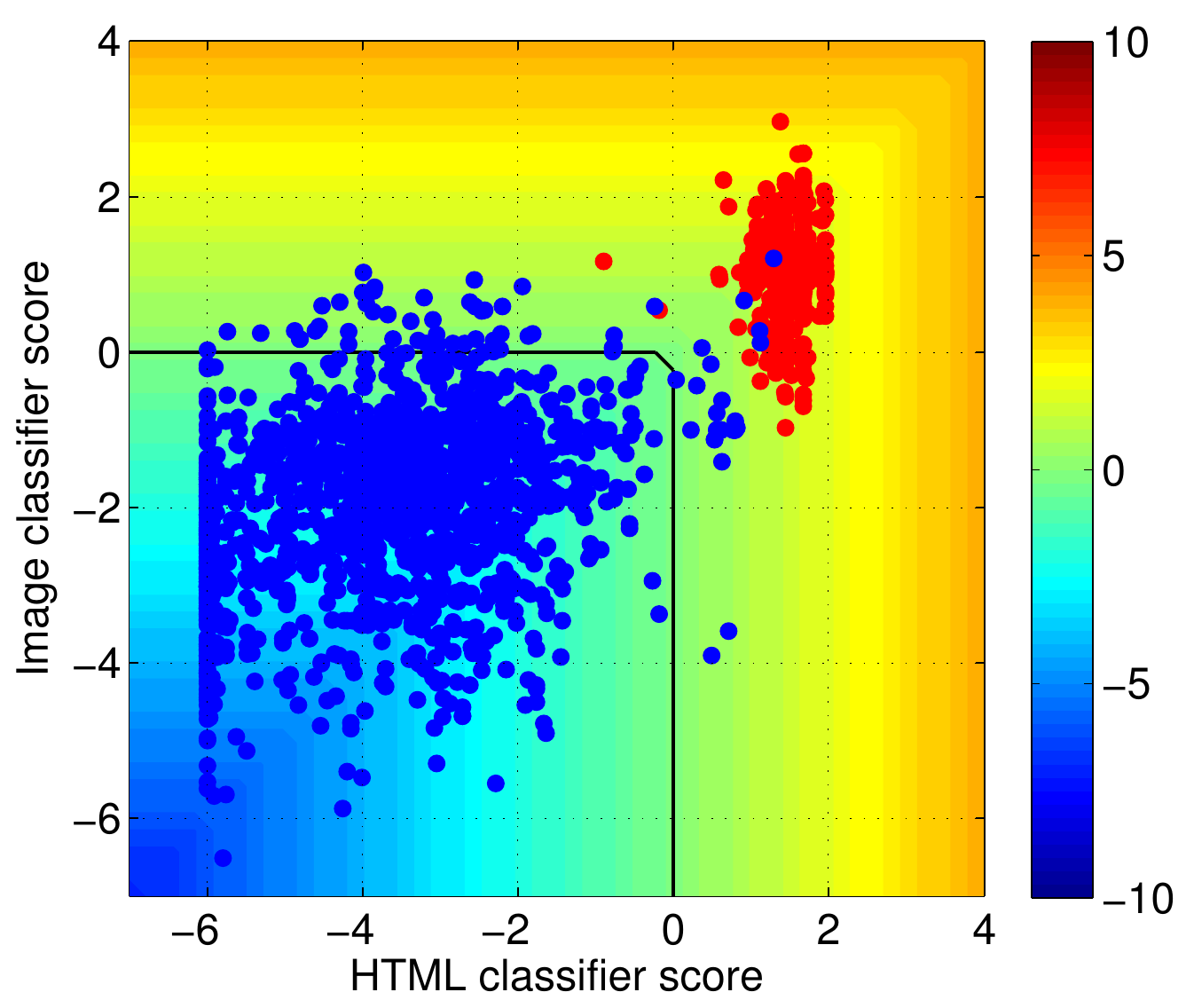}
\includegraphics[width=0.325\textwidth]{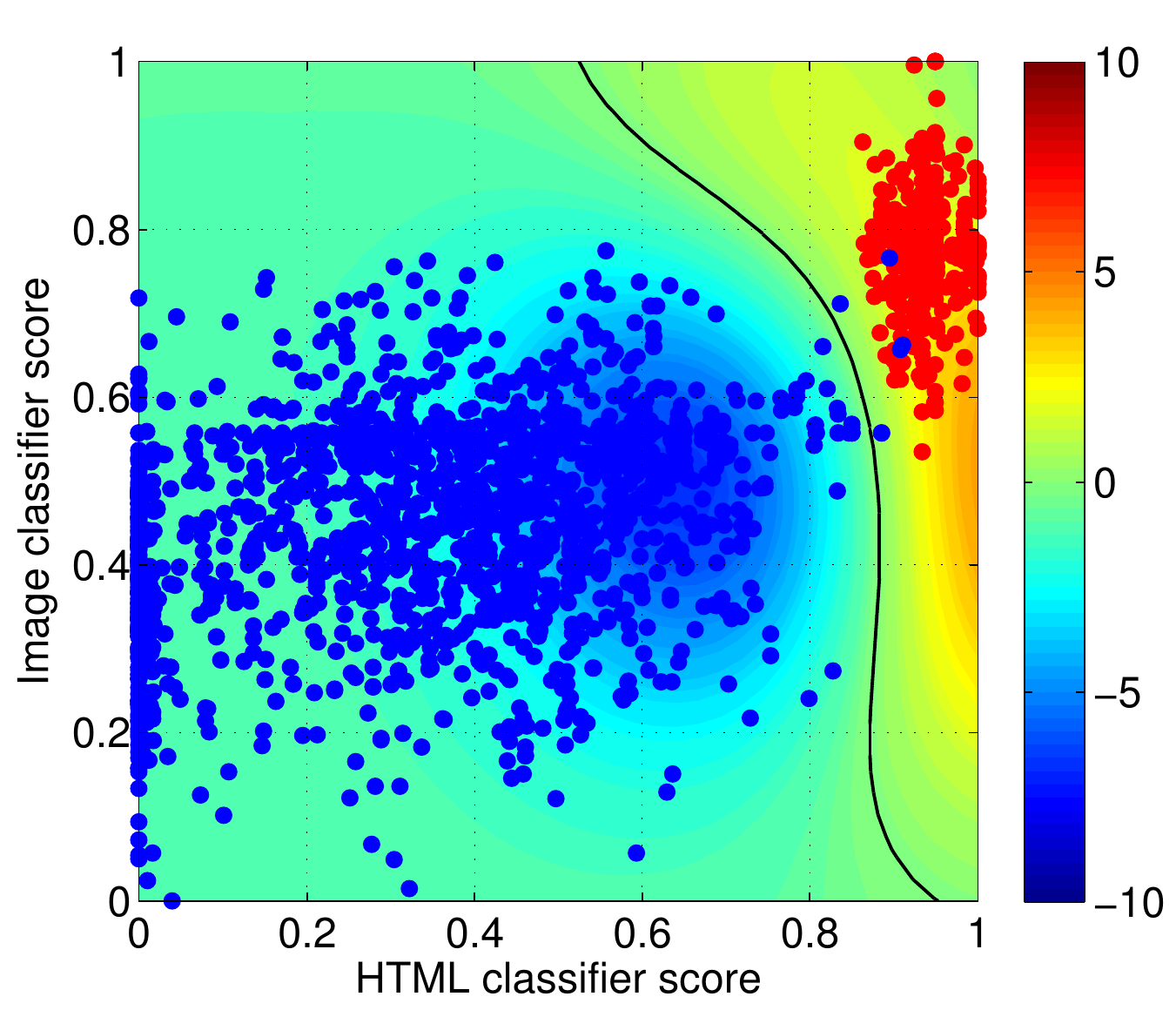}
\includegraphics[width=0.325\textwidth]{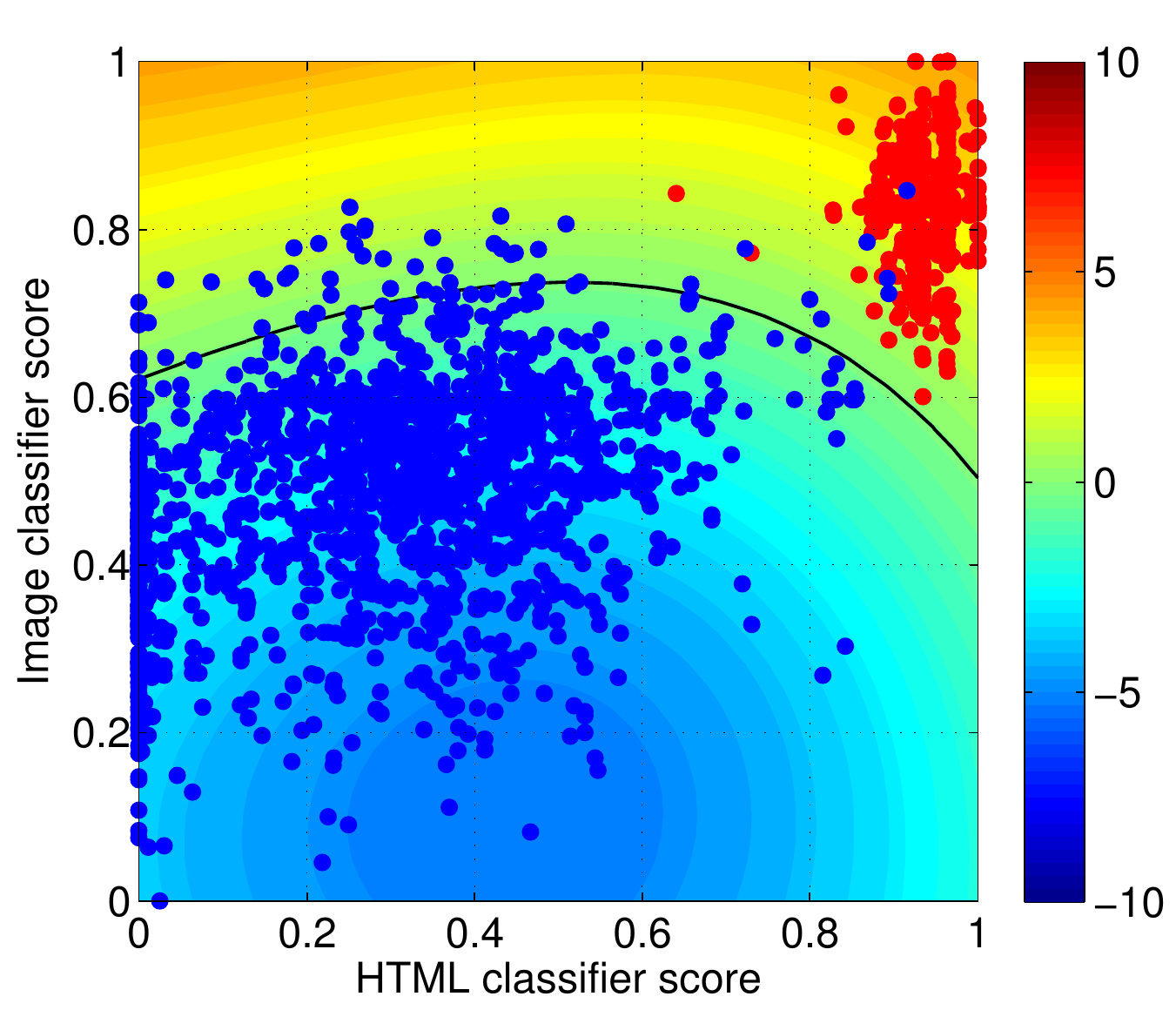}
\caption{Examples of decision functions (in colors) for maximum (\emph{left}), trained fusion (\emph{center}), and adversarial fusion (\emph{right}), in the space of the base classifiers' outputs. Blue (red) points represent legitimate (phishing) pages. Decision boundaries are shown as black lines. Phishing pages manipulated to evade the HTML-based classifier will receive a lower score (\ie, the red points will be shifted to the left), and most likely evade only the trained fusion.}
\label{fig:secure-fusion}
\end{figure*}

The HTML-based classifier is able to detect more than 97\% of phishing webpages while misclassifying less than 0.5\% of legitimate webpages, demonstrating the effectiveness of exploiting \emph{differences} in the HTML code of phishing and legitimate pages. The Snapshot-based classifier is not able to reach such accuracy since in some cases legitimate webpages may have some different visual appearance, and the visual learning task is inherently more complex. The visual classifier is indeed trained on a much higher number of features than the HTML-based one. Nevertheless, the detection rate of the Snapshot-based classifier is higher than 80\% at 1\% FP, which is still a significant achievement for this classification task.
Note finally that both trained and max fusion rules are able to achieve accuracy similar to those of the HTML-based classifier, while the adversarial fusion performs slightly worse.
This behavior is due to the fact that injecting simulated attacks into the training score set of the combiner causes an increase of the false positive rate (see Fig.~\ref{fig:secure-fusion}). This highlights a tradeoff between system security under attack and accuracy in the absence of targeted attacks against the HTML-based classifier.

\myparagraph{Processing time.} We have run our experiments on a personal computer equipped with an Intel(R) Xeon(R) CPU E5-2630 0 operating at  2.30GHz and 4 GB RAM.
The processing time of \deltaphish is clearly dominated by the browser automation module, which has to retrieve the HTML code and snapshot of the considered pages. This process typically requires few seconds (as estimated, on average, on our dataset). The subsequent HTML-based classification is instantaneous, while the Snapshot-based classifier requires more than 1.2 seconds, on average, to compute its similarity score. This delay is mainly due to the extraction of the HOG features, while the color features are extracted in less than 3 ms, on average. The processing time of our approach can be speeded up using parallel computation (\eg, through the implementation of a scalable application on a cloud computing service), and a caching mechanism to avoid re-classifying known pages. 

\myparagraph{Adversarial Evasion.} We consider here an attacker that manipulates the HTML code of his/her phishing page to resemble that of the homepage of the compromised website, aiming to evade detection by our HTML-based classifier.
We simulate a worst-case scenario in which the attacker has perfect knowledge of such a classifier, \ie, that he/she knows the weights assigned by the classifier to each HTML feature.
The idea of this evasion attack is to maximally decrease the classification score of the HTML module while manipulating the minimum number of features, as in~\cite{biggio13-ecml}. In this case, an optimal attack will start manipulating features having the highest absolute weight values. For simplicity, we assume a worst case attack, where the attacker can modify a feature value either to $0$ or $1$, although this may not be possible for all features without compromising the nature of the phishing scam. For instance,  in order to set the URL feature to $1$ (see Sect.~\ref{sub-sec:HTML-Based}), an attacker has to use exactly the same set of URLs present in the compromised website's homepage. This might require removing some links from the phishing page, compromising its malicious functionality. 

\begin{figure}[t]
\centering
\includegraphics[width=0.45\textwidth]{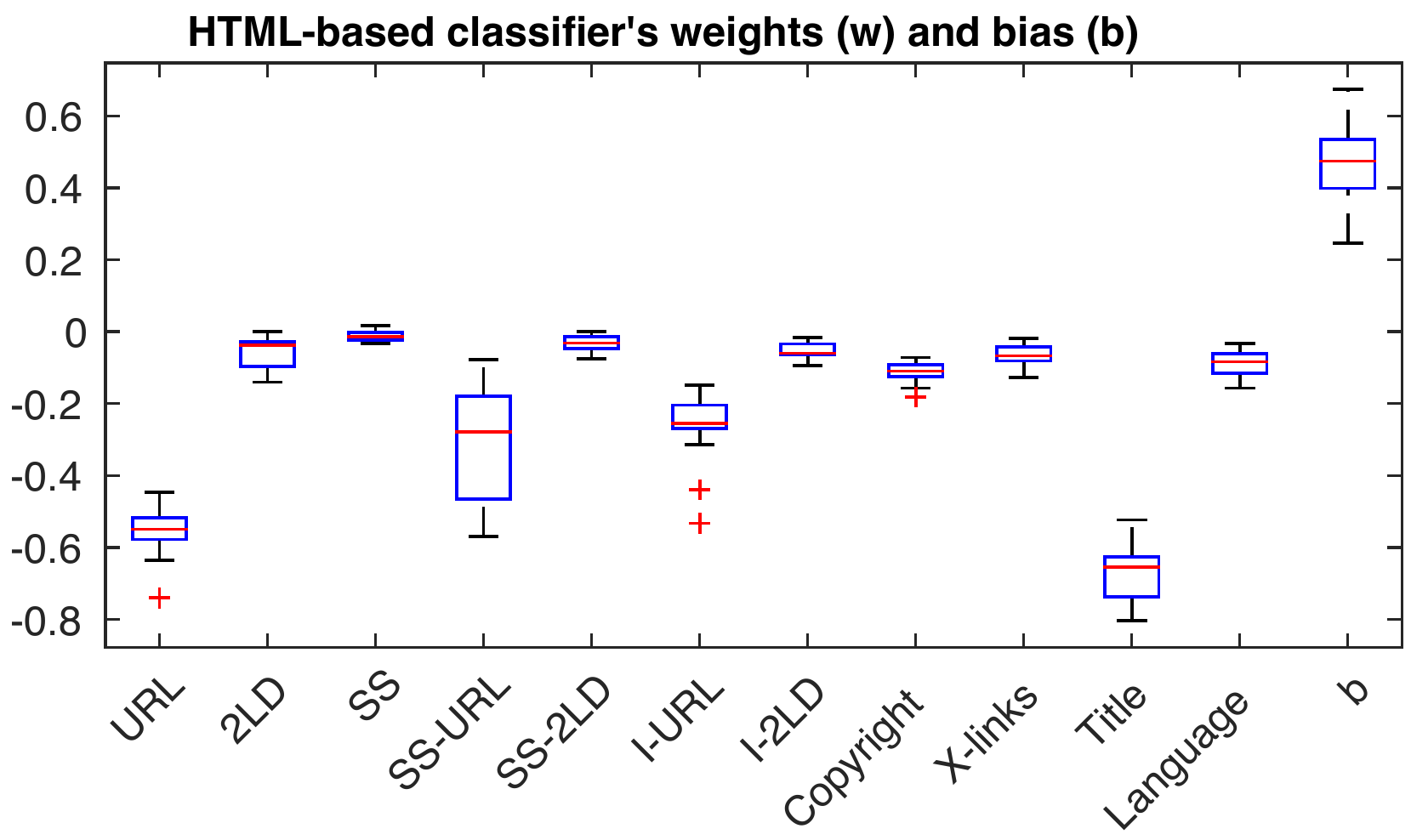}
\caption{Boxplot of feature weights (and bias) for the HTML-based classifier.}
\label{fig:boxplot}
\end{figure}

The distribution of the feature weights (and bias) for the HTML-based classifier (computed over the 20 repetitions of our experiment) is shown in the boxplot of Fig.~\ref{fig:boxplot}, highlighting two interesting facts. First, features tend to be assigned only negative weights. This means that each feature tends to exhibit higher values for legitimate pages, and that the attacker should increase its value to mislead detection.
Since the bias is generally positive, a page tends to be classified generally as a phish, unless there is sufficient ``evidence'' that it is similar to the homepage.
Second, the most relevant features (\ie, those which tend to be assigned the lowest negative weights) are \emph{Title}, \emph{URL}, \emph{SS-URL}, and \emph{I-URL}. This will be, in most of the cases, the first four features to be increased by the attacker to evade detection, while the remaining features play only a minor role in the classification of phishing and legitimate pages.

The results are reported in Fig.~\ref{fig:phish-det} (\emph{right} plot).
It shows how the detection rate achieved by \deltaphish at 1\% FP decreases against an increasing number of HTML features modified by the attacker, for the different fusion schemes and the HTML-based classifier.
The first interesting finding is about the HTML-based classifier, that can be evaded by modifying only a single feature (most likely, \emph{URL}).
The trained fusion remains slightly more robust, although it exhibits a dramatic performance drop already at the early stages of the attack.
Conversely, the detection rate of maximum and adversarial fusion rules under attack remains higher than 70\%. The underlying reason is that they rely more upon the output of the Snapshot-based classifier with respect to the trained fusion. In fact, as already mentioned, such schemes explicitly account for the presence of attacks against the base classifiers. Note also that the adversarial fusion outperforms maximum when only one feature is modified, while achieving a similar detection rate at the later stages of the attack. This clearly comes at the cost of a worse performance in the absence of attack. 
Thus, if one retains that such evasion attempts may be very likely in practice, he/she may decide to trade accuracy in the absence of attack for an improved level of security against these potential manipulations. This tradeoff can also be tuned in a more fine-grained manner by varying the percentage of simulated attacks while training the adversarial fusion scheme (which we set to 30\%), and also by considering a less pessimistic score distribution than the uniform one (\eg, a Beta distribution skewed towards the average score assigned by the HTML-based classifier to the phishing pages).

\vspace{-8pt}
\section{Conclusions and Future Work} \label{sect:conclusions}
\vspace{-5pt}
The widespread presence of public, exploitable websites in the wild
has enabled a large-scale deployment of modern phishing scams.
We have observed that phishing pages hosted in compromised websites exhibit a different aspect and structure from those of the legitimate pages hosted in the same website, for two main reasons: $(i)$ to be effective, phishing pages should resemble the visual appearance of the website targeted by the scam; and $(ii)$ leaving the legitimate pages intact guarantees that phishing pages remain active for a longer period of time before being blacklisted. Website compromise can be thus regarded as a simple \emph{pivoting} step in the implementation of modern phishing attacks.  

To the best of our knowledge, this is the first work that leverages this aspect for phishing webpage detection.
By comparing the HTML code and the visual appearance of a potential phishing page with the homepage of the corresponding website, \deltaphish exhibits high detection accuracy even in the presence of well-crafted, adversarial manipulation of HTML code.
While our results are encouraging, our proposal has its own limitations. It is clearly not able to detect phishing pages hosted through other means than compromised websites.
It may be adapted to address this issue by comparing the webpage to be classified against a set of known phishing targets (\eg, \texttt{PayPal}, \texttt{eBay}); in this case, if the similarity exceeds a given threshold, then the page is classified as a phish.
Another limitation is related to the assumption that legitimate pages within a certain website share a similar appearance/HTML code with the homepage. This assumption may be indeed violated, leading the system to misclassify some pages. We believe that such errors can be limited by extending the comparison between the potential phishing page and the website homepage also to the other legitimate pages in the website (and this can be  configured at the level of the web application firewall). 
This is an interesting evaluation for future work.

Our adversarial evaluation also exhibits some limitations. We have considered an attacker that deliberately modifies the HTML code of the phishing page to evade detection. A more advanced attacker might also modify the phishing page to evade our snapshot-based classifier. This is clearly more complex, as he/she should not compromise the visual appearance of the phishing page while aiming to evade our visual analysis. Moreover, the proposed adversarial fusion (\ie, the maximum) already accounts for this possibility, and the attack can be successful only if both the HTML and snapshot-based classifiers are fooled. We anyway leave a more detailed investigation of this aspect to future work, along with the possibility of training our system using only legitimate data, which would alleviate the burden of collecting a set of manually-labeled phishing webpages.

Finally, it is worth remarking that we have experimented on more than $5,500$ webpages collected in the wild, which we have also made publicly available for research reproducibility.
Despite this, it is clear that our data should be extended to include more examples of  phishing and legitimate webpages, hopefully through the help of other researchers, to get more reliable insights on the validity of phishing webpage detection approaches.

\subsection*{Acknowledgments}
This work has been partially supported by the DOGANA project, funded by the EU Horizon 2020 framework programme, under Grant Agreement no. 653618. 


\end{document}